\documentclass[twocolumn,superscriptaddress,amsmath,amssymb,showpacs,prl]{revtex4-2}

\usepackage{graphicx}
\usepackage{color}
\usepackage{makecell}
\usepackage{footnote}
\usepackage[colorlinks=true, allcolors=blue]{hyperref}
\renewcommand{\phi}{\varphi}

\begin{document}

\title{Spin state of \( \mathrm{Fe}^{2+} \) in \(I\overline{4}2d\)-type Mg\(_2\)SiO\(_4\) at ultra-high pressure}

\author{Tianqi Wan}
	\affiliation{Department of Applied Physics and Applied Mathematics, Columbia University, New York, NY 10027, USA}

\author{Yang Sun}
	\affiliation{Department of Applied Physics and Applied Mathematics, Columbia University, New York, NY 10027, USA}
	\affiliation{Department of Physics, Xiamen University, Xiamen 361005, China}

\author{Renata M. Wentzcovitch}
	\email{rmw2150@columbia.edu}
	\affiliation{Department of Applied Physics and Applied Mathematics, Columbia University, New York, NY 10027, USA}
	\affiliation{Department of Earth and Environmental Sciences, Columbia University, New York, NY 10027, USA}
	\affiliation{Lamont–Doherty Earth Observatory, Columbia University, Palisades, NY 10964, USA}

\date{\today}

\begin{abstract}

Under significantly higher pressures of approximately 500 GPa, typical of deep interiors of super-Earths, the combination of NaCl-type MgO and MgSiO$_3$ PPv have been reported to result in the formation of \(I\overline{4}2d\)-type Mg\(_2\)SiO\(_4\) (\textit{pppv}). This \textit{pppv} silicate phase could be the primary mantle silicate within these massive rocky exoplanets. Therefore, the fundamental properties of the \textit{pppv} phase, particularly in solid-solution with $\mathrm{Fe}_2\mathrm{Si}\mathrm{O}_4$ are of paramount significance. In this study, we present an \textit{ab initio} investigation on the properties of \( \mathrm{Fe}^{2+} \)-bearing \textit{pppv} from 400 GPa to 1 TPa. Given the localized nature of \textit{d}-electrons in iron, LDA + $U_{SC}$ and conventional DFT methods were used to investigate the electronic structure of this system. The dependence of U on volume and spin state are carefully considered in this system. We extensively explore the influence of pressure, temperature, and structure on the spin state of iron in the \( \mathrm{Fe}^{2+} \)-bearing \textit{pppv}, providing valuable information for modeling the mantle of super-Earth-type exoplanets. \\
\textbf{Keywords}: First Principles; Intermediate Spin State; Super-Earth; 

\end{abstract}

\maketitle

\section{I. Introduction}
More than 5500 exoplanets have been reported since the extraordinary discovery of a Jupiter mass planet around a sun-like star, Pegasi 51\cite{Mayor1995}. Among these exoplanets, super-Earths-type planets are arguably the most interesting. Investigating the distinctions between these exoplanets and our own Earth has become a vital pursuit, as it can deepen our understanding of the formation processes of planetary systems and offers insights into the search for habitable worlds. In this endeavor, researchers have initially approached the study by assuming compositions akin to Earth and other terrestrial planets within our solar system, gradually introducing complexity as our knowledge advances\cite{Doyle2019a}.

MgSiO$_3$ perovskite (Pv) is the major constituent of the Earth’s mantle and its highest-pressure polymorph in the mantle is post-perovskite (PPv)\cite{Murakami2004,Oganov2004,Tsuchiya2004}. In super-Earths, other forms of aggregation of MgO and SiO$_2$ stabilized by higher pressures and temperatures in their mantles are expected. Several post-PPv minerals like \(I\overline{4}2d\)-type Mg\(_2\)SiO\(_4\) and $P2_1/c$-type MgSi$_2$O$_5$, have been reported recently by \textit{ab initio} calculations\cite{Niu2015,Umemoto2011, Umemoto2006a, Umemoto2017a,Wu2013}. So far, these post-PPv phase transitions have not been confirmed experimentally due to the extremely high pressures. Also, the presence of iron affects several properties significantly, such as elastic and seismic properties\cite{Cobden2024,Wentzcovitch2009,Wu2013,Zhuang2024a} and electrical and thermal conductivites\cite{Badro2004,Katsura1998,Xu1998}. Here, we focus on the characterization of the \( \mathrm{Fe}^{2+} \)-bearing \(I\overline{4}2d\)-type Mg\(_2\)SiO\(_4\) at ultra-high pressure and offer predictions for novel high-pressure experiments. 

As reported in previous research\cite{Wentzcovitch2009}, iron’s spin state directly affects the mineral phase's transport, elastic, and rheological properties. The presence of localized 3\textit{d} electrons in iron requires methods beyond standard density functional theory (DFT) to address their strongly correlated nature\cite{Cococcioni2005a,Georges1996a,Lanata2019c,Mazin1997a,Shorikov2010a}. Among these, the widely adopted DFT+U method introduces the Hubbard correction to standard DFT calculations, enhancing the accuracy of results\cite{Yang18,Liechtenstein1995a}. However, the reliability of DFT+U outcomes critically hinges on the appropriate determination of the Hubbard parameter U, which should be derived self-consistently and be volume- and spin-state dependent \cite{Cococcioni2005a, Cococcioni2019a,Floris2020a,Hsu2010a,Kulik2006a,Sun2020a,Tsuchiya2006a,Wentzcovitch2009}. We study iron spin states in \( \mathrm{Fe}^{2+} \)-bearing \(I\overline{4}2d\)-type Mg\(_2\)SiO\(_4\) and corresponding electronic structure up to 1 TPa using LDA+$U_{SC}$ and conventional DFT methods. The dependence of U on pressure, volume, and spin-state is carefully considered in the system at ultra-high pressures. A local distortion in Low Spin (LS) state, which is important for its stabilization, has also been included. Also, we use the quasi-harmonic approximation (QHA)\cite{Wallace1972} to compute the vibrational free energy.

\section{II. Methods}
\subsection{2.1 \textit{Ab initio} calculations}
\textit{Ab initio} calculations are done with the \texttt{Quantum ESPRESSO} code\cite{Giannozzi2017, Giannozzi2009}. The local density approximation (LDA) and LDA+$U_{SC}$ calculations use Vanderbilt's ultra-soft pseudopotentials\cite{Vanderbilt1990} with valence electronic configurations $3\textit{s}^23\textit{p}^63\textit{d}^{6.5}4\textit{s}^14\textit{p}^0$, $3s^23p^1$, and $2s^22p^4$ for Fe, Si, and O, respectively. The pseudopotential for Mg was generated by von Barth-Car’s method using five configurations $3\textit{s}^23\textit{p}^0$, $3\textit{s}^13\textit{p}1$, $3\textit{s}^13\textit{p}^{0.5}3\textit{d}^{0.5}$, $3\textit{s}^13\textit{p}^{0.5}$, and $3\textit{s}^13\textit{d}^1$ with decreasing weights 1.5, 0.6 0.3 0.3, and 0.2, respectively. These pseudopotentials for Fe, Mg, Si, and O were generated, tested, and previously used in numerous works\cite{Tsuchiya2004,Umemoto2011, Umemoto2006a, Umemoto2017a, Umemoto2008}.  Plane-wave energy cutoffs are 100 Ry and 800 Ry for electronic wave functions and spin-charge density and potentials, respectively. When structural optimization, the irreducible Brillouin zone of the 28-atom cells is sampled by a 4×4×4 Monkhorst-Pack mesh\cite{Monkhorst1976}. Finer k-point grids (6×6×6) are used in the calculations of projected density of states and charge density. Effects of larger energy cutoff and k-point sampling on calculated properties are insignificant. The convergence thresholds are 0.01 \text{eV}/\text{\AA} for all components ($f_x$, $f_y$, and $f_z$) of all forces, including atomic forces and averaged forces in the supercell, and 1×$10^{-7}$ eV for the total energy of the supercell, 28 atoms in this work. 

In our LDA+$U_{SC}$ calculations, we apply the Hubbard correction\cite{Yang18} specifically to the Fe-3\textit{d} states. To compute the Hubbard parameter U, we utilize density-functional perturbation theory (DFPT)\cite{Yang27}. The convergence threshold for the response function is 1×$10^{-6}$  eV. We employed an automated iterative scheme to ensure the self-consistency of the $U_{SC}$ parameter and optimize the structure and spin state simultaneously\cite{Sun2020a}. Initially, we consider an empirical U value of 4.3 eV and compute the energies associated with all possible occupation matrices for the HS ($S = 2$), IS ($S = 1$), and LS ($S = 0$) states, resulting in a total of 65 possibilities. Among these, we select the electronic configuration, i.e., the occupation matrix with the lowest energy, for further structural optimization of lattice parameters and atomic positions. During the structural optimization, a new U parameter is recalculated. This process continues until mutual convergence of the structure\cite{Sun2020a}. U is achieved, with a convergence threshold of 0.01 eV for the U parameter and the above mentioned convergence criteria for structural optimizations\cite{Hsu2011,Tsuchiya2006a}.

Phonon calculations are performed in 224-atom supercells using the finite-displacement 
method with the \texttt{PHONOPY} code\cite{Togo2015a} and LDA+$U_{SC}$ forces obtained 
with the \texttt{Quantum ESPRESSO}. To obtain the vibrational density of states (VDoS), 
we employ a \textbf{q}-point mesh of 12×12×12. The vibrational contribution to the free 
energy is then calculated using the quasi-harmonic approximation (QHA)\cite{Wallace1972} 
with the \texttt{qha} code\cite{Qin2019}.

\subsection{2.2 Free energy calculations}
Since we expect high temperatures $>$ 4,000 K in the mantle of super-Earths, it is necessary to include the vibrational and electronic entropy to address the free energy calculations. Recently, it has been highlighted\cite{Wan2022,Zhuang2021a} that including the electronic entropy within the Mermin functional\cite{Mermin1965a,Wentzcovitch1992a} in a continuum of temperatures $T_{el}$ is important for calculating the thermodynamic properties. In this study, we perform the static calculation within a continuum of electronic temperatures, $T_{el}$. We sample electronic temperatures from 1,000 to 7,000 K with a spacing of 2,000 K, employing temperature interpolations. Combining the vibrational entropy, $S_{vib}$, obtained from phonon dispersion calculations, with the electronic entropy, we compute the Gibbs free energy for all three spin states utilizing the QHA. Obviously, $T_{el}$= T when the system is in thermodynamic equilibrium. A common expression for the free energy in this case:
\begin{equation}
F(V,T,T_{\text{el}}) = F_{\text{static}}(V,T_{\text{el}}) + F_{\text{vib}}(V,T,T_{\text{el}}), 
\end{equation}
where
\begin{equation}
F_{\text{static}}(V,T_{\text{el}}) = F_{\text{Mermin}}(V,T_{\text{el}})
\end{equation}
is the total Mermin free energy at volume \( V \). Here,
\begin{equation}
F_{\text{Mermin}}(V,T_{\text{el}}) = E_{\text{static}}(V,T_{\text{el}}) - T_{\text{el}} S_{\text{el}}(V,T_{\text{el}}), 
\end{equation}
where \( E_{\text{static}}(V,T_{\text{el}}) \) is the self-consistent energy with orbital occupancies
\begin{equation}
f_{ki}(V,T_{\text{el}}) = \frac{1}{\exp\left(\frac{\hbar\left(E_{\textbf{k}i} - E_F\right)}{k_B T_{\text{el}}}\right) + 1}, 
\end{equation}
with \( E_{\textbf{k}i} \) being the one-electron energy of an orbital with wavenumber \( \textbf{k} \) and band index \( i \), and \( E_F \) being the Fermi energy. The electronic entropy is
\begin{equation}
S_{\text{el}} = -k_B \sum_{\textbf{k},i} \left[(1 - f_{\textbf{k}i}) \ln(1 - f_{\textbf{k}i}) + f_{\textbf{k}i} \ln f_{\textbf{k}i} \right]. 
\end{equation}
The vibrational energy is
\begin{align}
&F_{\text{vib}}(V,T,T_{\text{el}}) = \frac{1}{2} \sum_{\textbf{q},s} \hbar \omega_{\textbf{q},s}(V, T_{\text{el}} = 0) \notag \\
&+ k_B T \sum_{\textbf{q},s} \left\{ \ln \left[ 1 - \exp \left( - \frac{\hbar \omega_{\textbf{q},s}(V,T_{\text{el}})}{k_B T} \right) \right] \right\}, 
\end{align}
where \( \omega_{\textbf{q},s}(V) \) is the vibrational frequency of noninteracting phonons with wavenumber \( \textbf{q} \) and polarization index \( s \).

\section{III. Results}
\subsection{3.1 HS and IS \(\mathrm{Fe}^{2+}\) in \( \mathrm{Fe}^{2+} \)-bearing \textit{pppv}}
\begin{figure}
	\includegraphics[width=\linewidth]{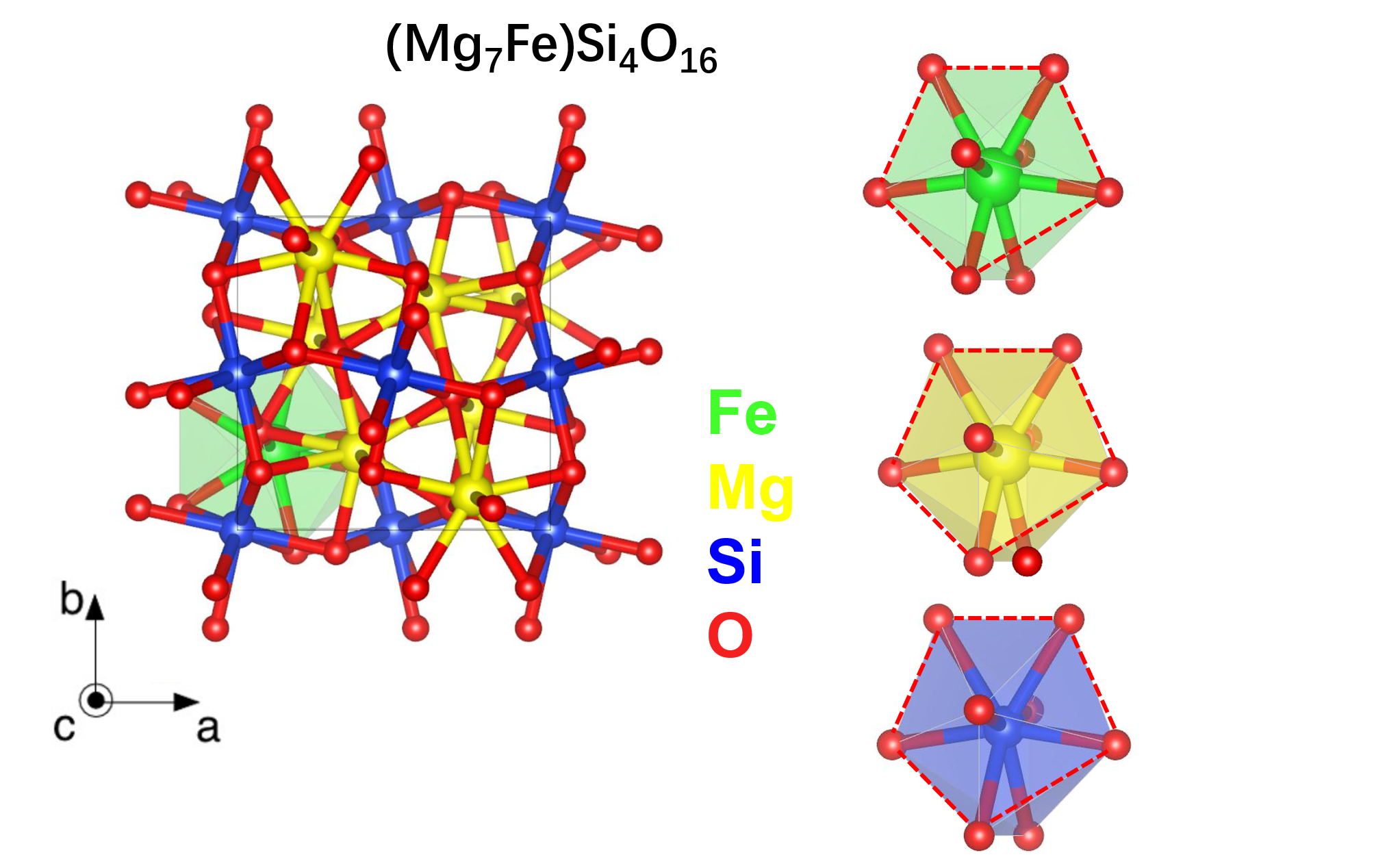}
	\caption{Crystal structure of \( \mathrm{Fe}^{2+} \)-bearing \(I\overline{4}2d\)-type Mg\(_2\)SiO\(_4\) at 1 TPa. Green, yellow, blue, and red spheres denote Fe, Mg, Si, and O ions.}
	\label{mg7fig1}
\end{figure}
The crystal structure of \(I\overline{4}2d\)-type Mg\(_2\)SiO\(_4\) is 
essential to understanding the electronic structure of the 
\(\mathrm{Fe}^{2+}\) ion. It exhibits a body-centered-tetragonal phase, which 
shares the same cation configuration as Zn$_2$SiO$_4$-II\cite{Marumo1971}. 
However, significant differences exist in the arrangement of oxygen atoms. Mg
\(_2\)SiO\(_4\) adopts a more densely packed structure compared to 
Zn$_2$SiO$_4$-II. The Zn and Si atoms in Zn$_2$SiO$_4$-II are tetrahedrally 
coordinated. In contrast, Mg and Si atoms in Mg\(_2\)SiO\(_4\) are coordinated 
by oxygens eightfold. MgO$_8$ and SiO$_8$ polyhedra exhibit striking 
similarities, with triangular faces forming pentagonal caps. The packing of 
Mg- and Si-centered polyhedra occurs through various edge- and face-sharing 
arrangements. Specifically, Si polyhedra share their edges, while Mg polyhedra 
share their faces. The crystal structures of \(I\overline{4}2d\)-type Mg
\(_2\)SiO\(_4\) together with the corresponding MO$_8$ (M=Mg, Fe, Si) 
polyhedra are shown in Fig. \ref{mg7fig1}. It is important to point out that 
the difference of average bond-lengths between MgO$_8$ and SiO$_8$ is also s
mall, shown in Table \ref{mg7table1}, 
\( \overline{\mathrm{Mg}-\mathrm{O}} = 1.6 \AA  \) and \( \overline{\mathrm{Si}-\mathrm{O}} = 1.57 \AA  \).

\begin{table}[ht]
\centering
\caption{Average bond length among different spin states at 1 TPa.}
\vspace{0.5em}
\begin{tabular}{|c|c|c|c|c|}
\hline
\textbf{Average Bond} & \textbf{HS} & \textbf{IS} & \textbf{Undistorted} & \textbf{Distorted} \\
\textbf{Length (Å)} &  &  & \textbf{LS} & \textbf{LS} \\
\hline
Fe-O & 1.640 & 1.641 & 1.637 & 1.640 \\
\hline
Mg-O & 1.60 & 1.60 & 1.60 & 1.60 \\
\hline
Si-O & 1.57 & 1.57 & 1.57 & 1.57 \\
\hline
\end{tabular}

\label{mg7table1}
\end{table}

In the context of \(\mathrm{Fe}^{2+}\) substitution within \(I\overline{4}2d\)-type Mg\(_2\)SiO\(_4\), our focus is primarily on the state of ferrous iron. However, it is important to acknowledge that ferric iron (\(\mathrm{Fe}^{3+}\)) can concurrently enter through coupled substitutions at the Mg and Si sites, as Mg and Si in this structure should be disordered\cite{Dutta2023, Dutta2022,Umemoto2021}. However, locally, we expect the \(\mathrm{Fe}^{2+}\) coordination be well described by the one adopted here. For \(\mathrm{Fe}^{2+}\) in the Mg-site, there can be HS ($d_{\uparrow}^5 d_{\downarrow}^1$), IS ($d_{\uparrow}^4 d_{\downarrow}^2$) state, and LS ($d_{\uparrow}^3 d_{\downarrow}^3$) states. Using the LDA+$U_{SC}$ method, all of these spin states have been carefully 
investigated. Fig. \ref{mg7fig5}(a) showcases the self-consistent Hubbard parameters. Notably, the HS state consistently exhibits lower U values than the LS states, a trend observed in ferropericlase\cite{Hsu2014a,Wan2022}. This trend is also found in FeO where the LS state of \(\mathrm{Fe}^{2+}\) always shows the largest self-consistent U value, regardless of crystal structure\cite{Sun2020a}. The $U_{SC}$ values exhibit variations of approximately 1 eV within the volume range inspected corresponding to 400 GPa < P < 1 TPa. Thus, in addition to the dependence on electronic configuration, the Hubbard parameter manifests a significant volume dependence. This phenomenon was initially observed in the study of the spin-crossover of \(\mathrm{Fe}^{2+}\) in ferropericlase\cite{Tsuchiya2006a}.

\begin{figure}
	\includegraphics[width=\linewidth]{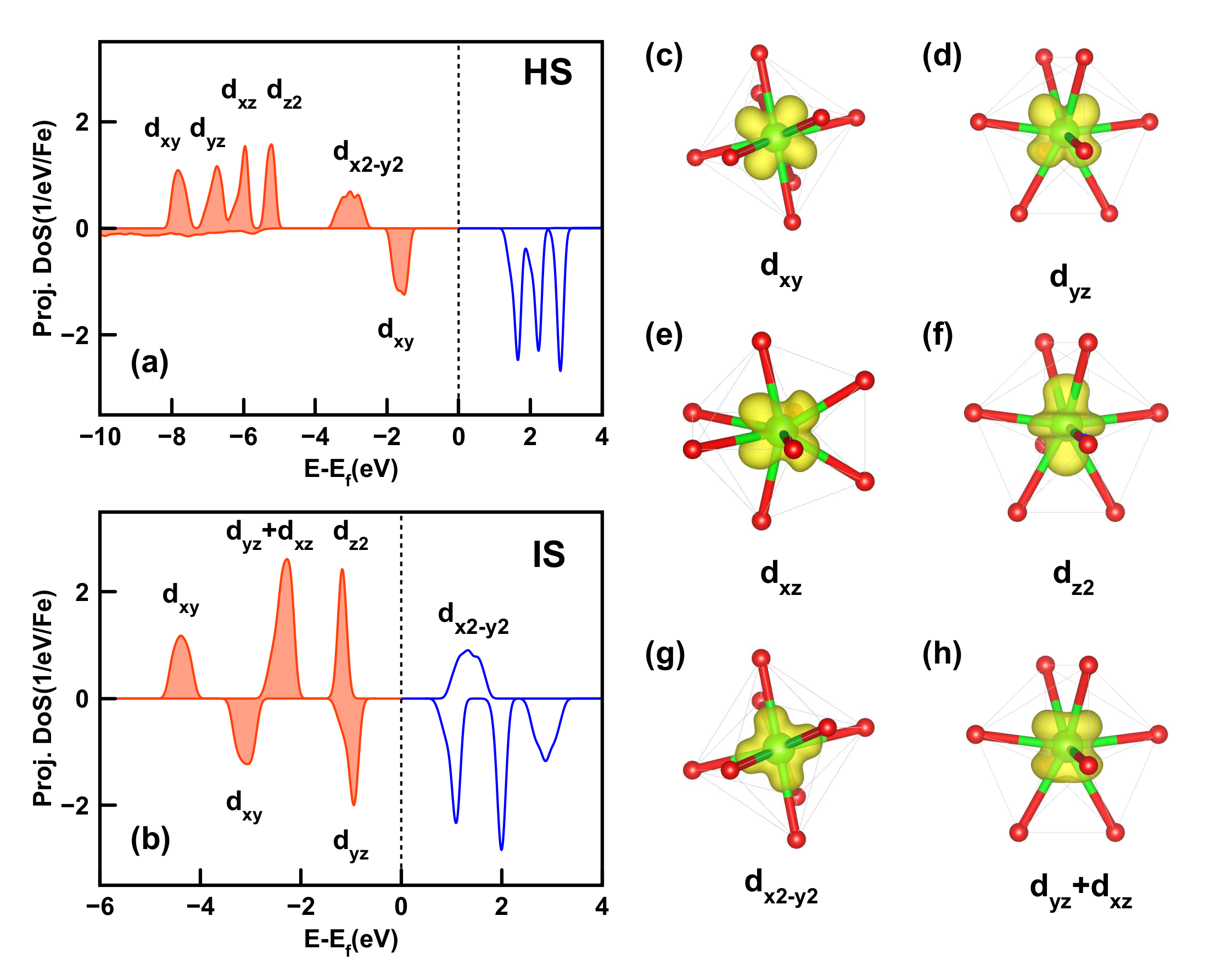}
	\caption{Projected density of states (PDOS) at 1 TPa for Fe 3\textit{d} orbitals in (a) HS and (b) IS. (c)-(g) present the charge density (yellow) for the different occupied orbitals in HS. (h) present the charge density (yellow) for the occupied doublet orbital in IS.}
	\label{mg7fig2}
\end{figure}

The projected density of states (PDOS) and charge density of HS and IS at 1 TPa are shown in Fig. \ref{mg7fig2}(a) and (b). In the HS state, all spin-up orbitals are fully occupied by five electrons, while the remaining electron occupies the spin-down $d_{xy}$ orbital, producing the $d_{\uparrow}^5 d_{\downarrow}^1$ electronic configuration with $S = 2$. It is important to note that the FeO$_8$ coordination in this case does not form a perfect cube. The Mg-Site configuration and Jahn-Teller distortion lead to the complete removal of the typical \textit{d}-level degeneracy, i.e., a triplet (the $t_{2g}$ states) and a doublet (the $e_g$ state) in octahedral or cubic environments, as seen in B1- and B2-ferropericlase\cite{Wan2022}. Instead, the five 3\textit{d} orbitals in this system become completely non-degenerate, forming five $a_{1g}$ singlets. As shown in Fig. \ref{mg7fig2}(c), the $d_{xy}$ orbital points away from the neighboring negatively charged oxygen atoms, while the $d_{x^2-y^2 }$ points toward them, thereby exhibiting the highest energy. In contrast, in the IS state, four electrons occupy the $d_{xy}$, $d_{xz}$, $d_{yz}$, and $d_{z^2}$ orbitals in the spin-up channel, while the remaining two electrons occupy the spin-down $d_{xy}$ and $d_{yz}$ orbitals. This leads to an electronic configuration of $d_{\uparrow}^4 d_{\downarrow}^2$ with a total spin $S = 1$. Additionally, as the spin state changes to the IS state, the energy of the $d_{yz}$ and $d_{xz}$ orbitals in the spin-up channel develop into a doublet, as shown in Fig. \ref{mg7fig2}(b) and (h).

\subsection{3.2 LS \(\mathrm{Fe}^{2+}\) in \( \mathrm{Fe}^{2+} \)-bearing \textit{pppv}}
\begin{figure}
	\includegraphics[width=\linewidth]{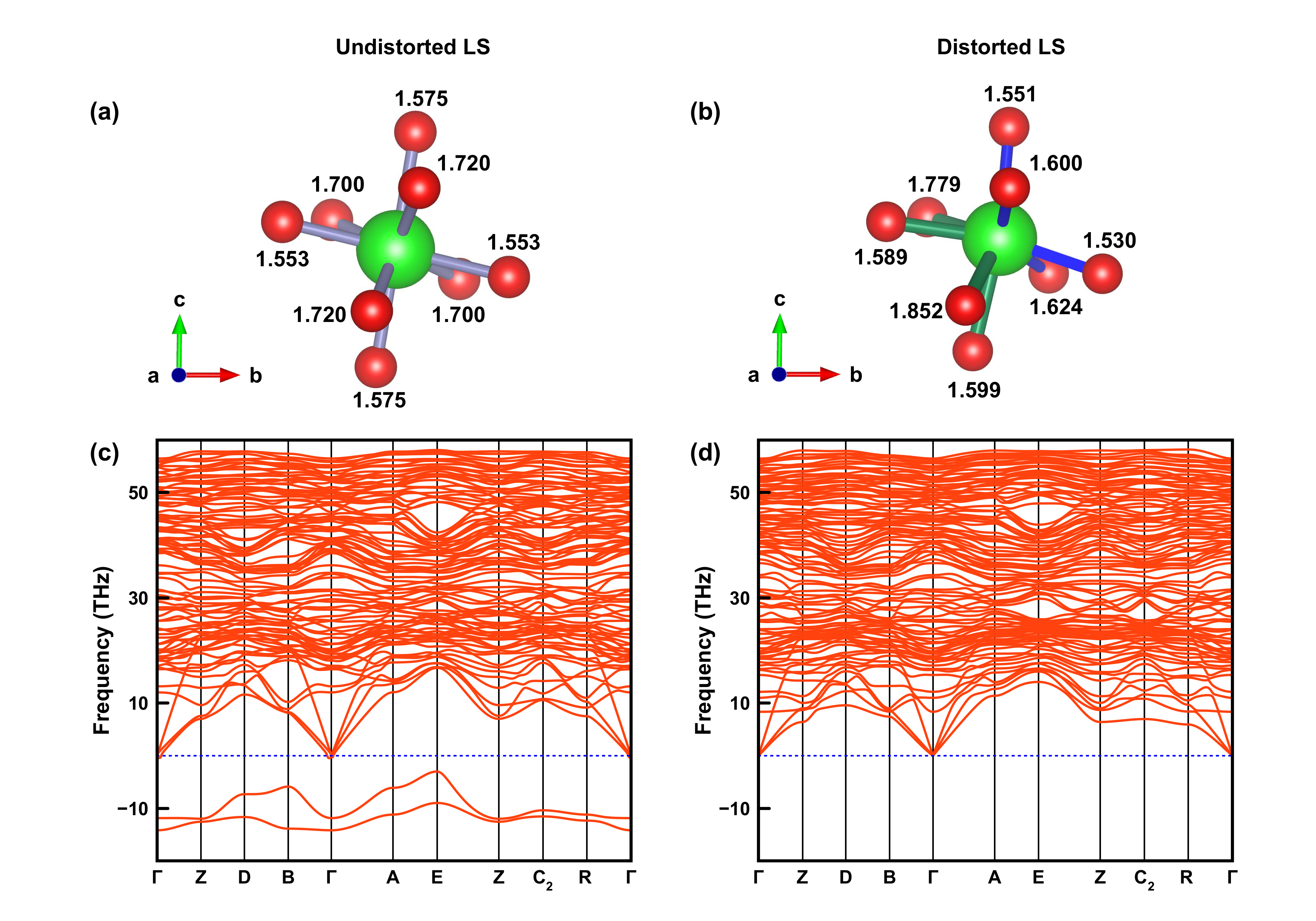}
	\caption{Local atomic configurations around iron and phonon dispersion at 1 TPa in (a)\&(c) undistorted LS state and (b)\&(d) distorted LS state. Numbers next to oxygens are Fe–O bond lengths (in \AA). The [100] direction is pointing out of the paper.}
	\label{mg7fig3}
\end{figure}

Furthermore, we have also considered the LS state. In the LS state, the electron occupying the $d_{z^2}$ orbital with spin up in the IS state can fill the $d_{xz}$ orbital with spin down, resulting in the LS state with an electronic configuration of $d_{\uparrow}^3 d_{\downarrow}^3$. However, this LS state exhibits imaginary phonon instabilities, as depicted in Fig. \ref{mg7fig3}(c). To address this issue, we introduce a displacement mode corresponding to the largest imaginary frequency around the gamma point and performed structural optimization after such displacement. Another displacement mode corresponding to the second largest imaginary frequency around the gamma point yields the same distorted structure. Fig. \ref{mg7fig3}(d) demonstrates that this newly distorted LS state is dynamically stable, as the phonon instabilities are no longer present.

The investigation of these two competing LS states deserves closer attention. The atomic structures of these two states show great resemblance, including the eight-fold coordination. The main difference between them lies in the iron position. Fig. \ref{mg7fig3}(a) and (b) shows the atomic arrangements around iron in these two states. By displacing iron along the [001] and [110] directions in the undistorted LS state, several Fe-O bond lengths change. In the undistorted LS state, there are four pairs of Fe-O band lengths which split upon such displacements. Specifically, four out of the eight bonds shorten and four lengthen. Notably, one of the two longest Fe-O bonds in the undistorted LS state shortens from 1.72 \AA to 1.6 \AA at 1 TPa. Table 1 shows an average decrease of approximately 0.2\% in the Fe-O bond length between the undistorted LS state and the HS or IS state. Such decrease in the Fe-O bond length, is no longer present after the distortion occurs. Therefore, the average Fe-O bond length in the distorted LS state is nearly identical to that in the HS or IS state.

\begin{figure}
\centering
	\includegraphics[width=\linewidth]{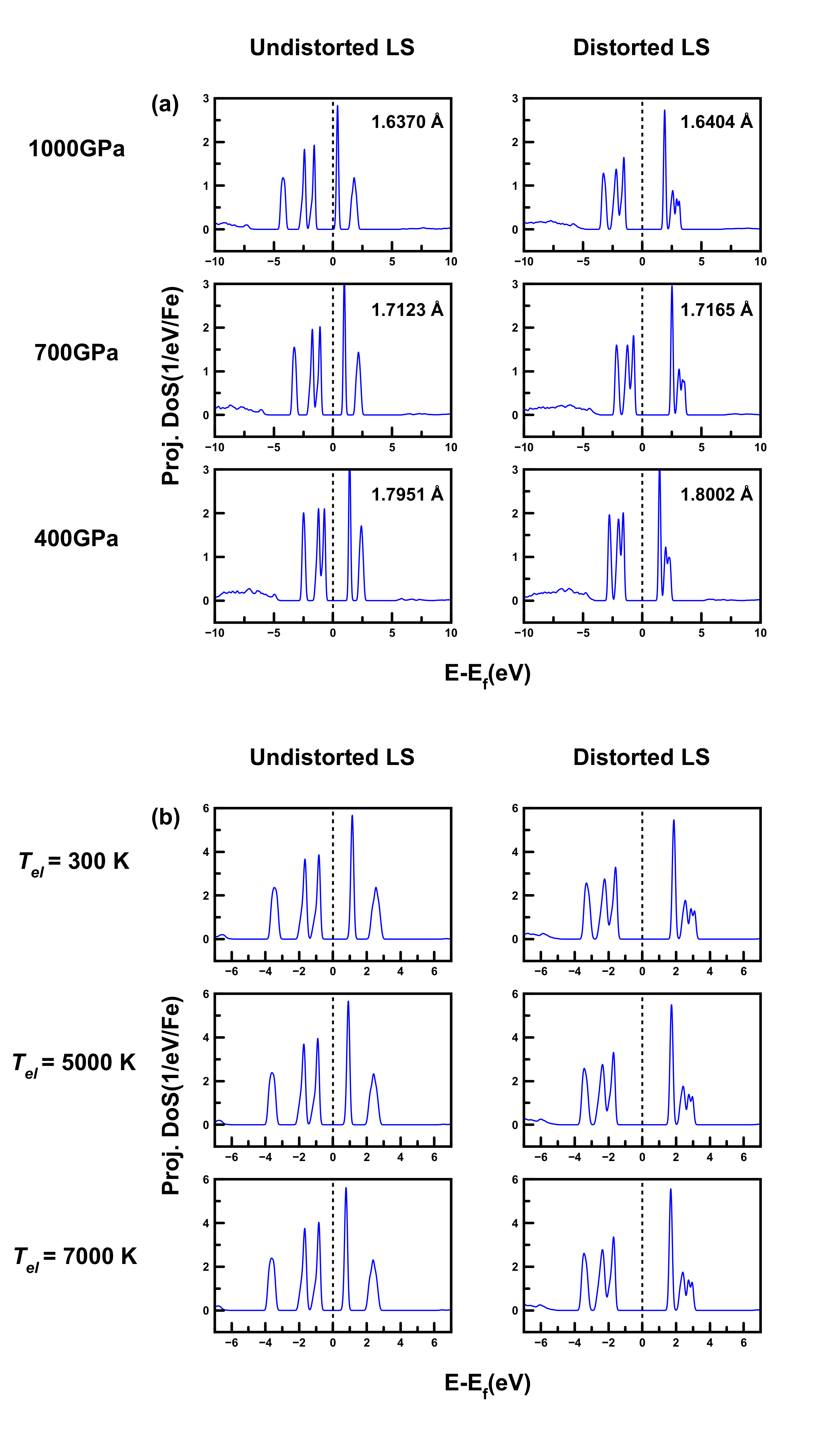}
	\caption{Projected density of states (PDOS) of two competing LS state at (a) different pressures; (b) different $T_{el}$.}
	\label{mg7fig4}
\end{figure}

Despite the different Fe-O arrangements, the \textit{d}-orbital occupancies of iron remain the same after the distortion. Fig. \ref{mg7fig4} shows the projected density of states (PDOS) of the two competing LS states at different pressure and $T_{el}$. We note that the undistorted LS state is an insulator with an energy gap slightly larger than 1.5 eV at 400 GPa. As the pressure increases, the energy gap gradually decreases and approaches $\sim$1 eV at 1 TPa, indicating a tendency towards metallicity at higher pressures. However, the band gap increases to around 2.7 eV at 400 GPa in the distorted structure, and the influence of pressure on the band gap becomes less pronounced. Furthermore, including the Mermin functional\cite{Mermin1965a,Wentzcovitch1992a} provides a first glimpse of the electronic structure at higher temperature which is more realistic for exoplanetary interiors. Similarly, the undistorted LS state remains insulating state with an energy gap of $\sim$1 eV at $T_{el}$  =7,000 K, and 1.5 eV at $T_{el}$  =300 K. The distortion enlarges the gap to $\sim$2.8 eV, which remains almost constant from $T_{el}$  =300 K to 7,000 K. Therefore, this distortion plays a crucial role in stabilizing not only the vibrational properties but also the insulating state against pressure and temperature.

\subsection{3.3 Stability of IS \(\mathrm{Fe}^{2+}\) in \( \mathrm{Fe}^{2+} \)-bearing \textit{pppv}}

\begin{figure}
	\includegraphics[width=\linewidth]{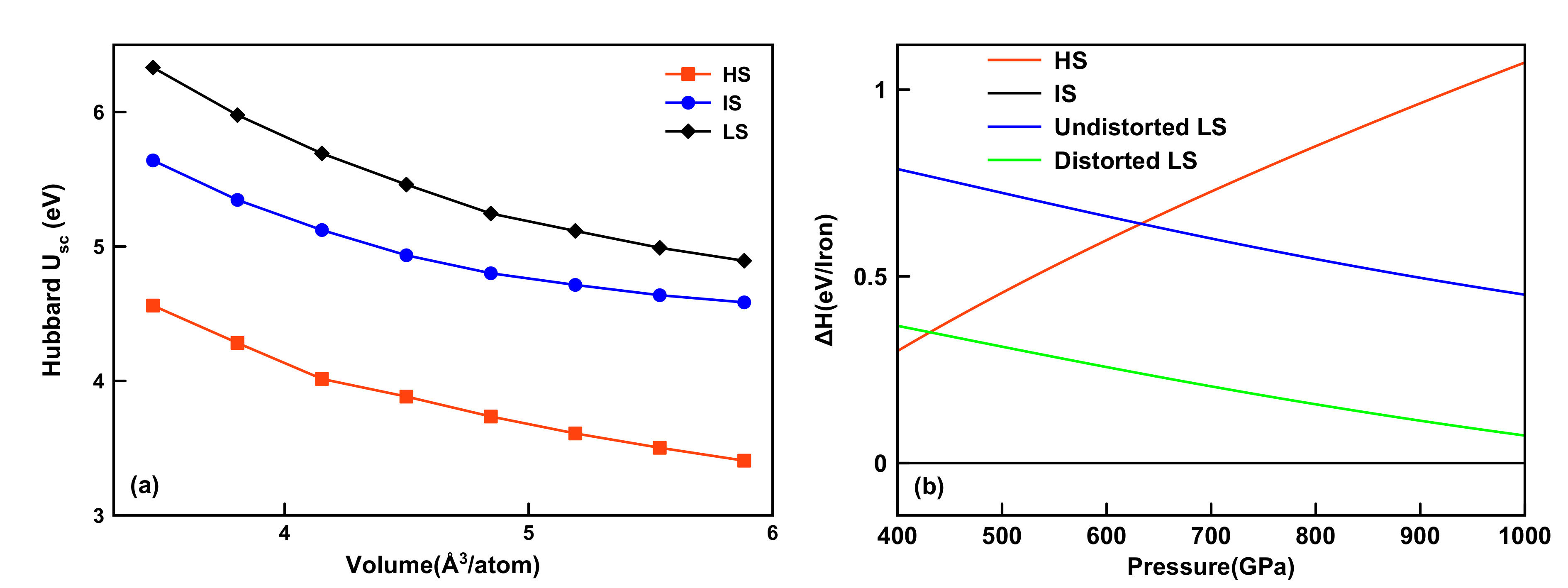}
	\caption{The self-consistent Hubbard parameters U vs volume. (b) Relative enthalpy ($\Delta H_i$) of \(\mathrm{Fe}^{2+}\) in \( \mathrm{Fe}^{2+} \)-bearing \(I\overline{4}2d\)-type Mg\(_2\)SiO\(_4\) in different spin state with respect to the IS state.}
	\label{mg7fig5}
\end{figure}

Using the LDA+$U_{SC}$ method, the enthalpies of \(\mathrm{Fe}^{2+}\)-bearing \textit{pppv} with all spin states can be computed. The energy-volume results of each state are fitted using the third-order Birch-Murnaghan (BM) equation of state (EoS). The relative enthalpy ($\Delta H_i$) of each spin state i [i=HS, IS, undistorted LS, and distorted LS] with respect to the IS state are plotted in Fig. \ref{mg7fig5}(b). The IS state consistently exhibits a lower enthalpy throughout the investigated pressure range, with no enthalpy crossing from HS to IS or IS to LS observed in this pressure range. Both LS states are energetically unfavorable, with distorted LS states being more stable than undistorted LS state, as expected.

\begin{figure}
	\includegraphics[width=\linewidth]{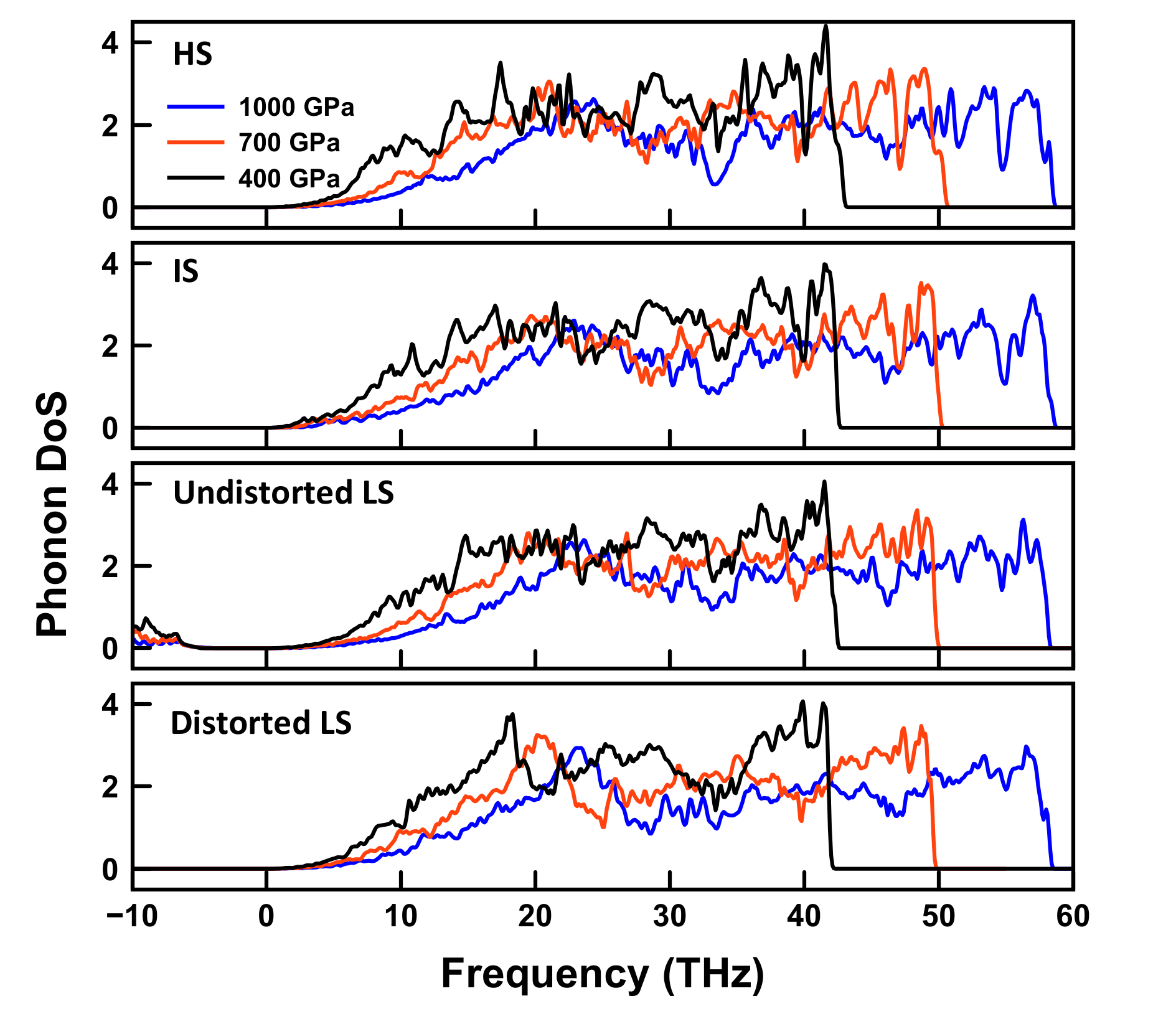}
	\caption{Phonon density of state for \( \mathrm{Fe}^{2+} \)-bearing \(I\overline{4}2d\)-type Mg\(_2\)SiO\(_4\) with different spin states.}
	\label{mg7fig6}
\end{figure}

\begin{figure}
	\includegraphics[width=\linewidth]{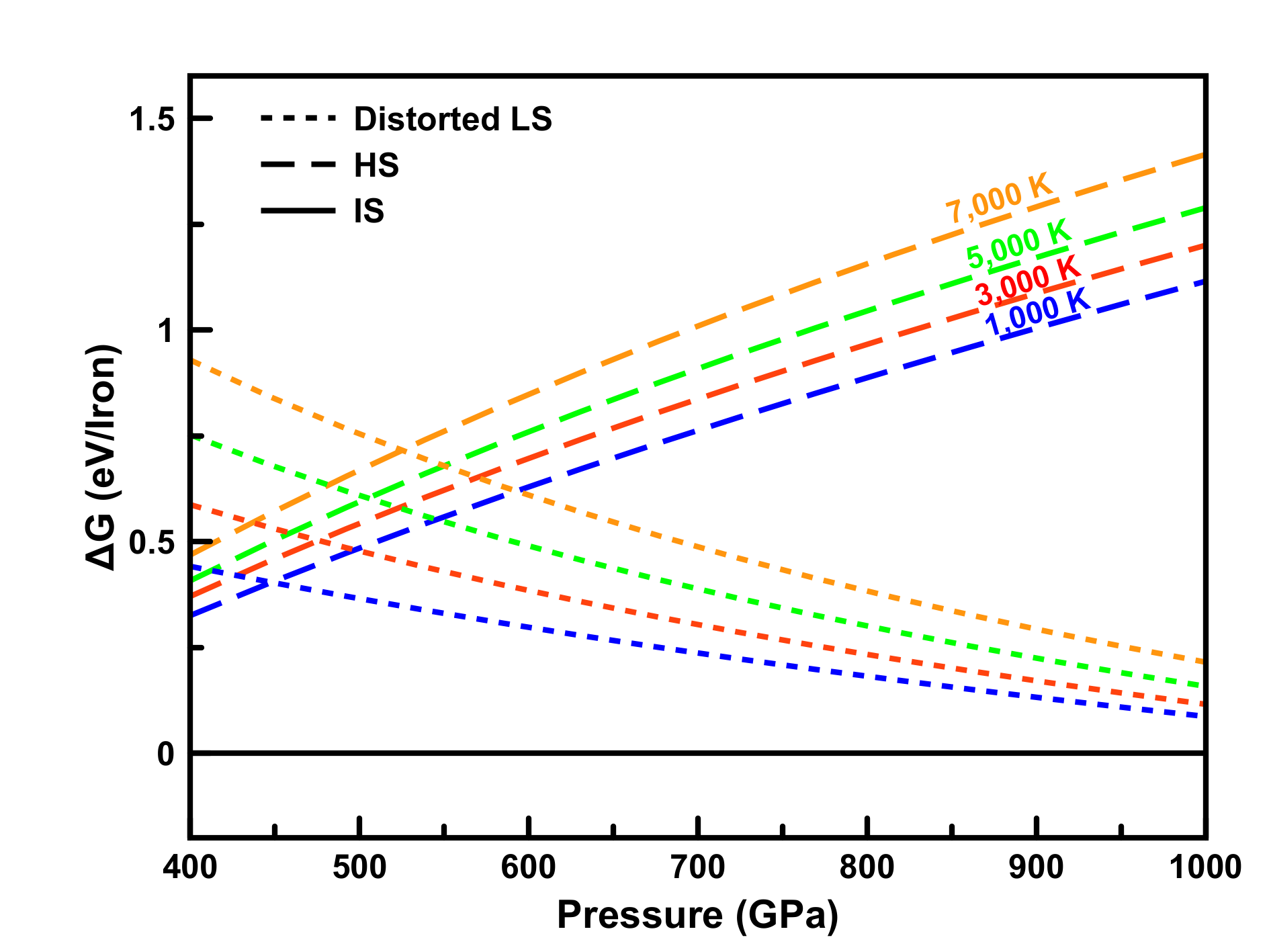}
	\caption{Relative Gibbs free energy ($\Delta G_i$) of \( \mathrm{Fe}^{2+} \)-bearing \(I\overline{4}2d\)-type Mg\(_2\)SiO\(_4\)  in each spin state with respect to the IS state. The black solid line corresponds to the IS state. Dot lines and dashed lines correspond to distorted LS state and HS state, respectively. Colors denote different temperatures.}
	\label{mg7fig7}
\end{figure}

We further analyze these states’ stability at high temperatures by computing the vibrational density of states and the vibrational free energy using \texttt{qha} code\cite{Qin2019}. Fig. \ref{mg7fig6} displays the phonon density of states of different spin states ranging from 400 GPa to 1 TPa. The undistorted LS state exhibits imaginary frequencies at all pressures, indicating the necessity of the distortion to stabilize phonons in the LS state. Furthermore, this stabilization is closely related to the stability of the insulating state. Increasing pressure increases phonon frequencies in HS, IS, and distorted LS states. No imaginary frequencies are observed in the investigated pressure. Therefore, the HS, IS, and distorted LS states of \(\mathrm{Fe}^{2+}\) in \( \mathrm{Fe}^{2+} \)-bearing \textit{pppv} with $x_{Fe}$=0.125 are dynamically stable at high temperatures. Fig. \ref{mg7fig7} presents the relative Gibbs free energy ($\Delta G_i$) of different spin states with respect to the IS state at various temperatures. We see that within the considered pressure range, the IS state remains the ground state even at 7,000 K. With increasing temperature, $\Delta G_{LS-IS}$ gradually increases, further stabilizing the IS state. On the other hand, pressure tends to stabilize the distorted LS state. In other words, pressure can potentially induce an IS to LS spin state change.

\section{IV. Conclusion}

We have investigated the stability of various spin states of \(\mathrm{Fe}^{2+}\) in \( \mathrm{Fe}^{2+} \)-bearing \(I\overline{4}2d\)-type Mg\(_2\)SiO\(_4\) at ultra-high pressures using LDA+$U_{SC}$ calculations. In the HS state, with $S=2$ and $d_{\uparrow}^5 d_{\downarrow}^1$ occupancy, we find all 3\textit{d} orbitals to be singlets with $a_{1g}$ symmetry. In the IS state with $S=2$ and $d_{\uparrow}^4 d_{\downarrow}^2$ occupancy, a doublet consisting of $d_{yz}$ and $d_{xz}$ orbitals appears in the majority-spin channel. The LS state with $S=0$ and $d_{\uparrow}^3 d_{\downarrow}^3$ occupancy exhibits imaginary phonon frequencies, indicating structural instability. It is necessary to further relax this LS state to stabilize vibrations and reduce the enthalpy of this state. Phonon calculations confirm the dynamic stability of all three possible spin states within the investigated pressure range. The IS state is the most stable throughout the entire pressure range investigated, with no observed transitions between HS or LS states, irrespective of pressure and temperature.
Our calculations shed light on the significant influence of strong 3\textit{d} electron localization and structural distortions on the stabilization of the spin state in the \( \mathrm{Fe}^{2+} \)-bearing \textit{pppv} system at ultra-high pressures and temperatures. However, given the complexity of the problem, it will be important to consider further electron-phonon and phonon-phonon interactions in the future, as they likely play important roles in determining the behavior of the ferrous ion. Additionally, the possibility of iron atoms substituting the Si$^{4+}$ site with a similar atomic environment in cation-disordered Mg$_2$SiO$_4$ and the inclusion of aluminum in this phase will greatly impact the electronic behavior of the system.

\section{Acknowledgments}
This research was primarily supported by the Gordon and Betty Moore Foundation Award GBMF \#12801, National Science Foundation Award No. EAR-1918126, and a Seed-Grant from the Center for Matter at Atomic Pressure, a National Science Foundation (NSF) Physics Frontier Center, under Award PHY-2020249. R.M.W. and Y.S., also acknowledge partial support from the Department of Energy, Theoretical Chemistry Program through Grant No. DE-SC0019759. Computational resources were provided by the Extreme Science and Engineering Discovery Environment funded by the National Science Foundation through Award No. ACI-1548562. The authors also acknowledge the Texas Advanced Computing Center at The University of Texas at Austin for providing high-performance computing resources that have contributed to the research results reported within this paper.

\twocolumngrid

\bibliographystyle{apsrev4-1}


\end{document}